\documentclass[amsmath,amssymb,preprint]{revtex4-1}
\usepackage{amsfonts} 
\usepackage{amsmath}
\usepackage{graphicx}
\usepackage{color}

\newcommand{\be}{\begin{eqnarray}} 
\newcommand{\ee}{\end{eqnarray}}
\newcommand{\D}{\mathrm{d}}
\renewcommand{\vec}{\mathbf}

\begin{document}

\title{Universal Poisson statistics of a passive tracer diffusing in dilute active suspensions}

\author{Adrian Baule$^1$}

\affiliation{
$^1$School of Mathematical Sciences, Queen Mary University of London, London E1 4NS, United Kingdom
}
\email{a.baule@qmul.ac.uk}

\begin{abstract}

The statistics of a passive tracer immersed in a suspension of active self-propelled particles (swimmers) is derived from first principles by considering a perturbative expansion of the tracer interaction with the microscopic swimmer field. To first order in the swimmer density, the tracer statistics is {\it exactly} represented as a spatial Poisson process combined with independent swimmer--tracer scattering events, rigorously reducing the multi-particle dynamics to two-body interactions. The Poisson representation is valid in any dimensions and for arbitrary interaction forces and swimmer dynamics. It provides in particular an analytical derivation of the coloured Poisson process introduced in [{\it K. Kanazawa et al.; Nature 579, 364 (2020)}] highlighting that such a non-Markovian process can be obtained from Markovian dynamics by a variable transformation.

\end{abstract}

\maketitle

A tracer immersed in a suspension of active swimmer particles such as self-propelled colloids or micro-organisms exhibits a range of statistical features that are different from Brownian motion, manifest, e.g., in an enhanced diffusion coefficient and a non-Gaussian displacement distribution \cite{Wu:2000aa,Leptos:2009aa,Kurtuldu:2011aa,Jeanneret:2016aa,Kurihara:2017aa}. Even though such features have first been observed more than two decades ago, there has so far been no comprehensive theory that is able to derive them from an analytical treatment of the multi-particle dynamics. Recent work has relied on mode-coupling approximations \cite{Reichert:2021aa,Reichert:2021ab}, specific interactions modelled as a weak linear force \cite{Maes:2020aa}, or is restricted to static swimmers \cite{Zaid:2016aa} or dynamics in one dimension \cite{Granek:2021aa}. In \cite{Kanazawa:2020aa} the tracer dynamics has been described as a coloured Poisson process predicting a L\'evy flight regime of the tracer due to the hydrodynamic interaction between the tracer and the swimmers. While such an approach successfully captures many empirical features of the tracer observed in simulations and experiments, it relies on a phenomenological assumption of independent scattering events and is only valid for a specific swimmer dynamics that leaves the random initial orientations constant in time. Here, I show that under very general assumptions an exact Poisson representation of the tracer dynamics can be derived when the swimmer density is low, which is essentially a manifestation of the discrete particle nature in the dilute regime.

The setup consists of $N$ active particles (swimmers) and a passive tracer suspended in a viscous fluid inside a box of volume $V$. In line with previous approaches \cite{Kanazawa:2020aa}, I assume: (a) a {\it dilute suspension} of swimmers with small number density $\rho_0=N/V$, such that mutual hydrodynamic interactions between the swimmers do not play a role; (b) a {\it passive tracer} that does not interact with the swimmer dynamics; (c) {\it weak thermal noise} that can be neglected compared with the active and viscous forces. The overdamped equation of motion for the tracer position $X(t)$ thus takes the form:
\be
\label{eqm}
\dot{\vec{X}}(t)=\mu\sum_{i=1}^N\vec{F}\left(\vec{Y}_i(t)-\vec{X}(t),\dot{\vec{Y}}_i(t)\right).
\ee
In Eq.~\eqref{eqm}, $\mu$ denotes the tracer's mobility coefficient and $\vec{F}$ the force between the tracer and the $i$th swimmer at position $\vec{Y}_i$, which can include both short-range and/or long-range interactions. Hydrodynamic interactions such as those generated by swimming microorganisms also depend on the swimming direction \cite{Leptos:2009aa}, indicated in Eq.~\eqref{eqm} by the dependence on $\dot{\vec{Y}}_i$. At this point, the swimmer dynamics $\vec{Y}_i$ is not specified, but is assumed to follow a stochastic process due to some intrinsic randomness in the swimmer motion. This could be due to random reorganization of the swimming direction and possibly the swimming speed. In particular, $\vec{Y}_i$ can be identified with the widely studied models of active brownian particles (ABP), run-and-tumble particles (RTP), and active Ornstein-Uhlenbeck particles (AOU) \cite{Fodor:2018aa}. The initial tracer position is set to the origin $\vec{X}(0)=\vec{0}$, while the initial positions of the swimmers are assumed uniform in the volume and their initial velocities are drawn from a distribution $p(\vec{v}_0)$. Below we focus on forces without $\dot{\vec{Y}}_i$ dependence to ease the notation. 

The central quantity containing the tracer statistics is the characteristic functional (CF) of the tracer velocity $\dot{\vec{X}}(t)$ defined as
\be
\label{cf}
\psi_\vec{\dot{X}}[\vec{k}]=\left<\exp\left\{i\int_{0}^t \vec{k}(s)\cdot\vec{\dot{X}}(s)\D s\right\}\right>,
\ee
which can be used to determine both the velocity statistics and the statistics of displacement increments like $\Delta \vec{X}=\vec{X}(\Delta t)-\vec{X}(0)$, which are typically measured in experiments. In order to determine the CF, we introduce the exact microscopic density field of the swimmers $\rho(\vec{x},t)=\sum_{i=1}^N\delta(\vec{x}-\vec{Y}_i(t))$,
so that Eq.~\eqref{eqm} can be likewise expressed as
\be
\label{eqm2}
\dot{\vec{X}}(t)=\mu\int\D\vec{x}\,\vec{F}(\vec{x}-\vec{X}(t))\rho(\vec{x},t).
\ee
The key step is then to express the tracer dynamics of Eq.~\eqref{eqm2} as a field theory using the Martin-Siggia-Rose-formalism (MSR) \cite{Martin:1973aa}. This implicitly requires to take the thermodynamic limit $V\to\infty, N\to\infty$, while keeping the swimmer density $\rho_0=$ const. Following standard MSR-procedure and enforcing the dynamics of Eq.~\eqref{eqm2} via delta-functions at every time step allows us to determine averages over observables $A$ of the tracer dynamics formally as \cite{Lefevre:2007aa}
\be
\label{pi}
\left<A[\vec{X}]\right>&=&\left<\left<A[\vec{X}]e^{i\mu\int_0^t\D s\int\D\vec{x}\,\vec{g}(s)\cdot\vec{F}(\vec{x}-\vec{X}(s))\rho(\vec{x},s)}\right>_0\right>
\ee
Here, the notation $\langle...\rangle_0$ is used for the average $\int\mathcal{D}\left[\frac{\vec{g}}{2\pi}\right]\int\mathcal{D}[\vec{X}] ...\delta(\vec{X}(0))e^{-i\int_0^t\D s\,\vec{g}(s)\cdot\vec{\dot{X}}(s)}$ and the remaining average is taken with respect to the swimmer dynamics incorporated in $\rho(\vec{x},t)$. The CF is obtained by applying Eq.~\eqref{pi} to the observable $A[\vec{X}]=e^{i\int_{0}^t\D u\, \vec{k}(u)\cdot\vec{\dot{X}}(u)}$. Expanding the exponential of Eq.~\eqref{pi} yields a perturbative expansion that contains at the $n$th order the $n$-point correlation function of $\rho(\vec{x},t)$. Crucially, in the absence of swimmer interactions, it can be shown that the $n$-point functions are given to first order in $\rho_0$ as (see Appendix Sec.~\ref{Sec:npoint})
\be
\label{npoint}
\left<\rho(\vec{x}_1,t_1)\cdots\rho(\vec{x}_n,t_n)\right>&=&\rho_0\int\D\vec{y}_0\left<\delta(\vec{x}_1-\vec{Y}(t_1))\cdots\delta(\vec{x}_n-\vec{Y}(t_n))\right>_{\vec{Y}(0)=\vec{y}_0},
\ee
i.e., are fully expressed by the self-correlations of a single swimmer conditional on the starting position $\vec{Y}(0)=\vec{y}_0$. With Eqs.~(\ref{cf},\ref{pi},\ref{npoint}) the CF has the series representation in the dilute regime
\be
\psi_\vec{\dot{X}}[\vec{k}]&\approx& \left<e^{i\int_{0}^t\D u\, \vec{k}(u)\cdot\vec{\dot{X}}(u)}\right>_0\nonumber\\
&+&\rho_0\int\D\vec{y}_0\left<\left<e^{i\int_{0}^t\D u\, \vec{k}(u)\cdot\vec{\dot{X}}(u)}\sum_{n=1}^\infty\frac{(i\mu)^n}{n!}\prod_{i=1}^n\int_0^t\D s_i\,\vec{g}(s_i)\cdot\vec{F}(\vec{Y}(s_i)-\vec{X}(s_i))\right>_0\right>_{\vec{Y}(0)=\vec{y}_0}\nonumber\\
&=&1+\rho_0\int\D\vec{y}_0\left<\sum_{n=1}^\infty\frac{i^n}{n!}\left(\int_0^t\D s\,\vec{k}(s)\cdot\vec{f}_\mu[\vec{Y}(s)]\right)^n\right>_{\vec{Y}(0)=\vec{y}_0}\label{series}
\ee
after evaluating explicitly the averages $\langle...\rangle_0$ in each term of the series. The technical details of the last step in Eq.~\eqref{series} are discussed in detail in Appendix Sec.~\ref{Sec:proof}. In Eq.~\eqref{series}, the functional $\vec{f}_\mu[\vec{Y}]$ is determined by the two-body interaction between the tracer and a single swimmer. Denoting by $\vec{X}^{\rm s}$ the solution of the two-body interaction ODE $\dot{\vec{X}}(t)=\mu\,\vec{F}(\vec{Y}(t)-\vec{X}(t))$, in general $\vec{X}^{\rm s}$ can be expressed as functional of $\vec{Y}$ and I define $\vec{f}_\mu[\vec{Y}]=\frac{\D}{\D t}\vec{X^{\rm s}}(t)$. For example, for small $\mu$ the solution $\vec{X}^{\rm s}$ can be approximated in terms of a Picard iteration (see Eq.~\eqref{ax:picard}) and expressed as a Taylor series in $\mu$ leading to the first and second order approximations
\be
\label{fapprox}
\vec{f}_{\mu^1}[\vec{Y}]=\mu\,\vec{F}(\vec{Y}(t)),\qquad \vec{f}_{\mu^2}[\vec{Y}]=\mu\,\vec{F}(\vec{Y}(t))+\mu^2\nabla_{\vec{x}}\vec{F}(\vec{Y}(t))\int_0^t\D s\,\vec{F}(\vec{Y}(s)).
\ee

With Eq.~\eqref{series}, a Poisson representation of the tracer CF in the dilute regime is obtained
\be
\label{main}
\ln\,\psi_\vec{\dot{X}}[\vec{k}]=\rho_0\int\D\vec{y}_0\left(\left<e^{i\int_{0}^t\D u\, \vec{k}(u)\cdot\vec{f}_\mu[\vec{Y}(u)]}\right>_{\vec{Y}(0)=\vec{y}_0}-1\right)
\ee
which is my main result. Eq.~\eqref{main} is exact up to first order in $\rho_0$ and implies that the tracer dynamics can be expressed as a stochastic equation in the form
\be
\label{spatialP}
\dot{\vec{X}}(t)=\sum_{\vec{y}_0\in \Phi}\vec{f}_\mu[\vec{Y}(t)],
\ee
where $\Phi$ represents a {\it spatial Poisson process} with intensity $\rho_0$: in every sub-volume $\tilde{V}$, the number of swimmers $\tilde{N}$ is distributed according to $P(\tilde{N}=m)=(\rho_0\tilde{V})^{m}e^{-\rho_0\tilde{V}}/m!$ and their positions $\vec{y}_0$ are uniform. The correspondence between Eqs.~\eqref{spatialP} and \eqref{main} can be established by substituting Eq.~\eqref{spatialP} in the definition of the CF, Eq.~\eqref{cf}, and evaluating the average, see Sec.~\ref{Sec:cf}. We thus see that at low densities the tracer moves due to independent scatterings between the tracer and individual swimmers whose starting positions are uniformly distributed in the volume. Remarkably, the simplification of the multi-particle dynamics to two-body interactions is an exact result of the calculation and relies on the fact that the $n$-point functions reduce exactly to single-particle self-correlations to lowest order in $\rho_0$ as expressed in Eq.~\eqref{npoint}. When $\vec{F}$ includes a dependence on the swimmer orientation, Eq.~\eqref{main} is valid in the same form with adapted $\vec{f}_\mu$, see Appendix Sec.~\ref{Sec:mapping} for further technical details.

Eq.~\eqref{main} represents the formal solution of the passive tracer problem in any dimensions and for arbitrary interaction forces and swimmer dynamics. While evaluating the conditional average in Eq.~\eqref{main} in general is very challenging, well-controlled approximations are obtained by truncating $\vec{f}_\mu$ at low orders in $\mu$.

Considering the second order approximation $\vec{f}_{\mu^2}[\vec{Y}]$, Eq.~\eqref{fapprox}, there is no formalism to my knowledge that would allow the calculation of the average $\left<e^{i\int_{0}^t\D u\, \vec{k}(u)\cdot\vec{f}_{\mu^2}[\vec{Y}(u)]}\right>_{\vec{Y}(0)=\vec{y}_0}$ for a stochastic process $\vec{Y}$. However, progress can be made by simplifying the swimmer dynamics to $\vec{Y}(t)=\vec{y}_0+v_{\rm A}\vec{\hat{n}}\,t$, i.e., the swimmer is moving in a straight line with constant speed $v_{\rm A}$, whereby the initial position and direction $\vec{\hat{n}}$ is chosen randomly. For this special case, the average can be calculated in closed form even for complicated interaction forces such as the hydrodynamic dipol interaction due to the far-flow field (see Eq.~\eqref{ax:hydrof}) \cite{Leptos:2009aa}. In fact, in this case the resulting CF is equivalent to the one obtained in the coloured Poisson framework developed in \cite{Kanazawa:2020aa}, which can be shown by virtue of a variable transformation that converts the memory-less spatial Poisson process of Eq.~\eqref{main} into a temporal one including memory, see Appendix Sec.~\ref{Sec:mapping}. In this way, Eq.~\eqref{main} thus provides a rigorous foundation for the theory of \cite{Kanazawa:2020aa}, which captures many of the striking empirical observations of the tracer dynamics such as loopy trajectories, enhanced diffusion, and non-Gaussian power-law tails in the displacement distribution \cite{Kanazawa:2020aa}.

\begin{acknowledgements}

I am grateful for helpful discussions with K. Kanazawa, A. Cairoli, and R. Jack.

\end{acknowledgements}

\begin{appendix}

\section{Correlation functions of the density field}
\label{Sec:npoint}

In order to derive Eq.~\eqref{npoint}, we first determine the 2-point density correlation function for finite $N,V$. Using the definition of the microscopic density
\be
\rho(\vec{x},t)=\sum_{i=1}^N\delta(\vec{x}-\vec{Y}_i(t))
\ee
and the fact that the swimmers are non-interacting, we obtain
\be
\left<\rho(\vec{x}_1,t_1)\rho(\vec{x}_2,t_2)\right>&=&\sum_{i=1}^N\sum_{j=1}^N\left<\delta(\vec{x}_1-\vec{Y}_i(t_1))\delta(\vec{x}_2-\vec{Y}_j(t_2))\right>\nonumber\\
&=&N\left<\delta(\vec{x}_1-\vec{Y}(t_1))\delta(\vec{x}_2-\vec{Y}(t_2))\right>\nonumber\\
&&+N(N-1)\left<\delta(\vec{x}_1-\vec{Y}(t_1))\right>\left<\delta(\vec{x}_2-\vec{Y}(t_2))\right>.
\ee
The average is taken with respect to the initial positions, which are assumed uniform in the volume $V$, and the stochastic process underlying the dynamics of $\vec{Y}$. In the second term, the average over the initial positions yields 
\be
\label{ax:1point}
\left<\delta(\vec{x}_1-\vec{Y}(t_1))\right>&=&\frac{1}{V}\int \D \vec{y}_0\left<\delta(\vec{x}_1-\vec{Y}(t_1))\right>_{\vec{Y}(0)=\vec{y}_0}\nonumber\\
&=&\frac{1}{V}\int \D \vec{y}_0\,G(\vec{x}_1-\vec{y}_0,t_1)\nonumber\\
&=&\frac{1}{V}.
\ee
which is a simple consequence of the translation invariance of the propagator $G(\vec{x}_1-\vec{y}_0,t_1)=\left<\delta(\vec{x}_1-\vec{Y}(t_1))\right>_{\vec{Y}(0)=\vec{y}_0}$. As a consequence
\be
\left<\rho(\vec{x}_1,t_1)\rho(\vec{x}_2,t_2)\right>&=&\frac{N}{V}\int\D \vec{y}_0\left<\delta(\vec{x}_1-\vec{Y}(t_1))\delta(\vec{x}_2-\vec{Y}(t_2))\right>_{\vec{Y}(0)=\vec{y}_0}+\frac{N(N-1)}{V^2}
\ee
and introducing swimmer number density $\rho_0=N/V$ leads to the result in the thermodynamic limit
\be
\left<\rho(\vec{x}_1,t_1)\rho(\vec{x}_2,t_2)\right>&=&\rho_0\int\D \vec{y}_0\left<\delta(\vec{x}_1-\vec{Y}(t_1))\delta(\vec{x}_2-\vec{Y}(t_2))\right>_{\vec{Y}(0)=\vec{y}_0}+\rho_0^2.
\ee

For the 3-point correlation function we obtain likewise
\be
\left<\rho(\vec{x}_1,t_1)\rho(\vec{x}_2,t_2)\rho(\vec{x}_3,t_3)\right>&=&\sum_{i=1}^N\sum_{j=1}^N\sum_{l=1}^N\left<\delta(\vec{x}_1-\vec{Y}_i(t_1))\delta(\vec{x}_2-\vec{Y}_j(t_2))\delta(\vec{x}_3-\vec{Y}_l(t_3))\right>\nonumber\\
&=&N\left<\delta(\vec{x}_1-\vec{Y}(t_1))\delta(\vec{x}_2-\vec{Y}(t_2))\delta(\vec{x}_3-\vec{Y}(t_3))\right>\nonumber\\
&&+N(N-1)\left<\delta(\vec{x}_1-\vec{Y}(t_1))\right>\left<\delta(\vec{x}_2-\vec{Y}(t_2))\delta(\vec{x}_3-\vec{Y}(t_3))\right>\nonumber\\
&&+N(N-1)\left<\delta(\vec{x}_2-\vec{Y}(t_2))\right>\left<\delta(\vec{x}_1-\vec{Y}(t_3))\delta(\vec{x}_3-\vec{Y}(t_3))\right>\nonumber\\
&&+N(N-1)\left<\delta(\vec{x}_3-\vec{Y}(t_3))\right>\left<\delta(\vec{x}_2-\vec{Y}(t_2))\delta(\vec{x}_1-\vec{Y}(t_1))\right>\nonumber\\
&&+N(N-1)(N-2)\left<\delta(\vec{x}_1-\vec{Y}(t_1))\right>\left<\delta(\vec{x}_2-\vec{Y}(t_2))\right>\times\nonumber\\
&&\left<\delta(\vec{x}_3-\vec{Y}(t_3))\right>
\ee
Taking the thermodynamic limit and considering Eq.~\eqref{ax:1point} leads to
\be
\left<\rho(\vec{x}_1,t_1)\rho(\vec{x}_2,t_2)\rho(\vec{x}_3,t_3)\right>&=&\rho_0\int \D y_0\left<\delta(\vec{x}_1-\vec{Y}(t_1))\delta(\vec{x}_2-\vec{Y}(t_2))\delta(\vec{x}_3-\vec{Y}(t_3))\right>_{\vec{Y}(0)=\vec{y}_0}\nonumber\\
&&+\rho_0^2\left\{\int\D \vec{y}_0\left<\delta(\vec{x}_2-\vec{Y}(t_2))\delta(\vec{x}_3-\vec{Y}(t_3))\right>_{\vec{Y}(0)=\vec{y}_0}\right.\nonumber\\
&&+\int\D \vec{y}_0\left<\delta(\vec{x}_1-\vec{Y}(t_3))\delta(\vec{x}_3-\vec{Y}(t_3))\right>_{\vec{Y}(0)=\vec{y}_0}\nonumber\\
&&+\left.\int\D \vec{y}_0\left<\delta(\vec{x}_2-\vec{Y}(t_2))\delta(\vec{x}_1-\vec{Y}(t_1))\right>_{\vec{Y}(0)=\vec{y}_0}\right\}+\rho^3_0
\ee

It is then easy to see that for the $n$-point function the term to lowest order in $\rho_0$ contains the $n$-point position PDF of a single swimmer conditioned on $\vec{Y}(0)=\vec{y}_0$, i.e., contains only single-particle self-correlations. This is also the only fully connected contribution to the $n$-point function, while all higher orders in $\rho_0$ are at least partially disconnected. To first order in $\rho_0$ the $n$-point function can thus be approximated as in Eq.~\eqref{npoint}.

\section{Evaluating the averages in the series expansion}
\label{Sec:proof}

The main result Eq.~\eqref{main} relies on the evaluation of the averages $\langle...\rangle_0$ in the expansion Eq.~\eqref{series}, which yields the identity
\be
&&\left<e^{i\int_{0}^t\D u\, \vec{k}(u)\cdot\vec{\dot{X}}(u)}\sum_{n=1}^\infty\frac{(i\mu)^n}{n!}\prod_{i=1}^n\int_0^t\D s_i\,\vec{g}(s_i)\cdot\vec{F}(\vec{Y}(s_i)-\vec{X}(s_i))\right>_0\nonumber\\
&&=\sum_{m=1}^\infty\frac{i^m}{m!}\left(\int_0^t\D s\,\vec{k}(s)\cdot\vec{f}_\mu[\vec{Y}(s)]\right)^m,\label{ax:series}
\ee
where we define
\be
\left<\vphantom{\frac{\delta}{\delta}}...\right>_0=\int\mathcal{D}\left[\frac{\vec{g}}{2\pi}\right]\int\mathcal{D}[\vec{X}]...\delta(\vec{X}(0))e^{i\int_0^t\D s\,\vec{g}(s)\cdot\dot{\vec{X}}(s)}.
\ee
and the functional $\vec{f}_\mu[\vec{Y}]$ is obtained from the two-body interaction between the tracer and a single swimmer following trajectory $\vec{Y}$
\be
\label{ax:2body}
\dot{\vec{X}}(t)=\mu\,\vec{F}(\vec{Y}(t)-\vec{X}(t)).
\ee
If we consider $\vec{Y}$ as a given function prescribing the time dependence in the force and denote the solution of Eq.~\eqref{ax:2body} as $\vec{X}^{\rm s}$, we obtain $\vec{f}_\mu[\vec{Y}]$ as
\be
\label{ax:fmu}
\vec{f}_\mu[\vec{Y}]=\frac{\D}{\D t}\vec{X^{\rm s}}(t).
\ee
An explicit expression for $\vec{f}_\mu[\vec{Y}]$ can be derived from a fixed-point (Picard) iteration considering Eq.~\eqref{ax:2body} in integrated form
\be
\label{ax:picard}
\vec{X}(t)=\mu\int_0^t\D s\,\vec{F}(\vec{Y}(s)-\vec{X}(s)),
\ee
where $\vec{X}(0)=\vec{0}$. Up to second order we have then
\be
\vec{X}^{s,2}(t)&=&\mu\int_0^t\D s\,\vec{F}\left(\vec{Y}(s)-\mu\int_0^s\D s'\,\vec{F}\left(\vec{Y}(s')\right)\right)
\ee
and thus
\be
\vec{f}_\mu[\vec{Y}]\approx\mu\,\vec{F}\left(\vec{Y}(t)-\mu\int_0^t\D s\,\vec{F}\left(\vec{Y}(s)\right)\right).
\ee
Taylor expansion in orders of $\mu$ then yields the approximations in Eq.~\eqref{fapprox}. Before we establish Eq.~\eqref{ax:series} by evaluating the averages order-by-order in $\mu$, we present a heuristic argument for its validity.

\subsection{Heuristic proof based on the MSR-formalism}

Note that the lhs of Eq.~\eqref{ax:series} can be written as
\be
&&\left<e^{i\int_{0}^t\D u\, \vec{k}(u)\cdot\vec{\dot{X}}(u)}\sum_{n=1}^\infty\frac{(i\mu)^n}{n!}\prod_{i=1}^n\int_0^t\D s_i\,\vec{g}(s_i)\cdot\vec{F}(\vec{Y}(s_i)-\vec{X}(s_i))\right>_0\nonumber\\
&&=\left<e^{i\int_{0}^t\D u\, \vec{k}(u)\cdot\vec{\dot{X}}(u)}\left(e^{i\mu\int_0^t\D s\,\vec{g}(s)\cdot\vec{F}(\vec{Y}(s)-\vec{X}(s))}-1\right)\right>_0\nonumber\\
&&=\left<e^{i\int_{0}^t\D u\, \vec{k}(u)\cdot\vec{\dot{X}}(u)+i\mu\int_0^t\D s\,\vec{g}(s)\cdot\vec{F}(\vec{Y}(s)-\vec{X}(s))}\right>_0-1\label{ax:series2}
\ee
since $\langle...\rangle_0$ simply constrains the dynamics of $\vec{X}(t)$ to the constant motion $\dot{\vec{X}}(t)=0$ with initial tracer position $\vec{X}(0)=\vec{0}$ and thus $\left<e^{i\int_{0}^t\D u\, \vec{k}(u)\cdot\vec{\dot{X}}(u)}\right>_0=1$. The remaining average
\be
&&\left<e^{i\int_{0}^t\D u\, \vec{k}(u)\cdot\vec{\dot{X}}(u)+i\mu\int_0^t\D s\,\vec{g}(s)\cdot\vec{F}(\vec{Y}(s)-\vec{X}(s))}\right>_0\nonumber\\
&&=\int\mathcal{D}\left[\frac{\vec{g}}{2\pi}\right]\int\mathcal{D}[\vec{X}]e^{i\int_{0}^t\D u\, \vec{k}(u)\cdot\vec{\dot{X}}(u)+i\mu\int_0^t\D s\,\vec{g}(s)\cdot\vec{F}(\vec{Y}(s)-\vec{X}(s))-i\int_0^t\D s\,\vec{g}(s)\cdot\vec{\dot{X}}(s)}
\ee
is nothing but the MSR expression for the average $\left<e^{i\int_{0}^t\D u\, \vec{k}(u)\cdot\vec{\dot{X}}(u)}\right>$ constraining $\vec{X}(t)$ to the solution of the ODE Eq.~\eqref{ax:2body}. This implies that 
\be
\left<e^{i\int_{0}^t\D u\, \vec{k}(u)\cdot\vec{\dot{X}}(u)+i\mu\int_0^t\D s\,\vec{g}(s)\cdot\vec{F}(\vec{Y}(s)-\vec{X}(s))}\right>_0=e^{i\int_0^t\D u\,\vec{k}(u)\cdot\vec{f}_\mu[\vec{Y}(u)]}
\ee
using Eq.~\eqref{ax:fmu} and Eq.~\eqref{ax:series} follows.

\subsection{Proof based on term-by-term evaluation of averages}

In this section, we simplify notation by considering only one dimension and consider the general formulation of the identity Eq.~\eqref{ax:series} as a connection between the averages over $\langle...\rangle_0$ and the solution of an ODE. In general terms we consider the ODE
\be
\label{ax:ode}
\dot{X}(t)=\mu \,F(t,X(t)),\qquad\qquad X(0)=x_0,
\ee
where the explicit time-dependence in $F$ stems in our original problem from the dependence on the swimmer trajectory $Y(t)$. We assume that the solution $X^{\rm s}(t)$ of Eq.~\eqref{ax:ode} can be expressed as a power-series in $\mu$
\be
\label{ax:expansion}
X^{\rm s}(t)=x_0+\sum_{n=1}^\infty\mu^n\int_0^t\D s\,\phi_n(s,x_0).
\ee
The equivalent identity to Eq.~\eqref{ax:series} is then
\be
&&\left<e^{i\int_{0}^t\D u\, z(u)\dot{X}(u)}\sum_{n=1}^\infty\frac{(i\mu)^n}{n!}\prod_{i=1}^n\int_0^t\D s_i\,g(s_i)F(s_i,X(s_i))\right>_0\nonumber\\
&&=\sum_{m=1}^\infty\frac{i^m}{m!}\left(\int_0^t\D s\,z(s)\sum_{n=1}^\infty\mu^n\phi_n(s,x_0)\right)^m,\label{ax:identity}
\ee
which we will establish by considering order-by-order in $\mu$.

\subsubsection{The functions $\phi_n$}

One way to obtain a series representation like in Eq.~\eqref{ax:expansion} is to use the Picard iteration and then perform a Taylor expansion around $x_0$. For the first three orders, e.g., one requires the third Picard iteration
\be
X^{\rm s,3}(t)=\mu\int_0^t\D s\,F\left(s,x_0+\mu\int_0^s\D s'\,F\left(s',x_0+\mu\int_0^{s'}\D s''\,F(s'',x_0)\right)\right)
\ee
and the base functions $\phi_n$ in Eq.~\eqref{ax:expansion} are determined as
\be
\phi_1(s,x_0)&=&F(s,x_0)\label{ax:phi0}\\
\phi_2(s,x_0)&=&\partial_xF(s,x_0)\int_0^s\D s'\,F(s',x_0)\\
\phi_3(s,x_0)&=&\partial_xF(s,x_0)\int_0^s\D s'\,\partial_xF(s',x_0)\int_0^{s'}\D s''\,F(s'',x_0)\nonumber\\
&&+\frac{1}{2}\partial^2_xF(s,x_0)\int_0^s\D s'\,F(s',x_0)\int_0^{s}\D s''\,F(s'',x_0)
\ee

Crucially, the $n$th order function $\phi_n$ can likewise be obtained as a path integral average in the following way
\be
\label{ax:result}
\phi_n(s_n,x_0)&=&\frac{1}{(n-1)!}\int_0^t\D s_1...\int_0^t\D s_{n-1}\left<\frac{\delta}{\delta \dot{X}(s_1)}\cdots \frac{\delta}{\delta \dot{X}(s_{n-1})}F(s_1,X(s_1))\cdots\right.\nonumber\\
&&\left. \vphantom{\frac{\delta}{\delta}}F(s_{n-1},X(s_{n-1}))F(s_{n},X(s_{n}))\right>_0,
\ee
where the brackets again indicate an average with respect to paths constrained to $X(t)=x_0$. We verify the validity of Eq.~\eqref{ax:result} order by order noting that
\be
\label{ax:cancel}
\frac{\delta}{\delta \dot{X}(s_j)}F(s_l,X(s_l))=\int_0^{s_l}\D u\,\delta(u-s_j)\partial_xF(s_l,X(s_l))=0
\ee
when $j=l$ (due to the standard path integral discretization used) or when $s_l<s_j$. We then obtain
\be
\phi_1(s,x_0)&=&\left<F(s,X(s))\right>_0=F(s,x_0)\\
\phi_2(s_2,x_0)&=&\int_0^t\D s_1\left<\frac{\delta}{\delta \dot{X}(s_1)}F(s_1,X(s_1))F(s_2,X(s_2))\right>_0\nonumber\\
&=&\int_0^t\D s_1\Theta(s_2-s_1)\left<F(s_1,X(s_1))\partial_xF(s_2,X(s_2))\right>_0\nonumber\\
&=&\int_0^{s_2}\D s_1F(s_1,x_0)\partial_xF(s_2,x_0)\\
\phi_3(s_3,x_0)&=&\frac{1}{2}\int_0^t\D s_1\int_0^t\D s_2\left<\frac{\delta}{\delta \dot{X}(s_1)}\frac{\delta}{\delta \dot{X}(s_2)}F(s_1,X(s_1))F(s_2,X(s_2)F(s_3,X(s_3)))\right>_0\nonumber\\
&=&\frac{1}{2}\int_0^t\D s_1\int_0^t\D s_2\left\{\vphantom{\frac{\delta}{\delta}}\Theta(s_3-s_2)\Theta(s_3-s_1)\right.\times\nonumber\\&&\left<F(s_1,X(s_1))F(s_2,X(s_2)\partial^2_xF(s_3,X(s_3)))\right>_0\nonumber\\
&&+\Theta(s_3-s_2)\Theta(s_2-s_1)\left<F(s_1,X(s_1))\partial_xF(s_2,X(s_2)\partial_xF(s_3,X(s_3))\right>_0\nonumber\\
&&+\left.\Theta(s_3-s_1)\Theta(s_1-s_2)\left<\partial_xF(s_1,X(s_1))F(s_2,X(s_2)\partial_xF(s_3,X(s_3)))\right>_0\vphantom{\frac{\delta}{\delta}}\right\}\nonumber\\
&=&\frac{1}{2}\int_0^{s_3}\D s_2\int_0^{s_3}\D s_1F(s_1,x_0)F(s_2,x_0)\partial^2_xF(s_3,x_0)\nonumber\\
&&+\int_0^{s_3}\D s_2\int_0^{s_2}\D s_1F(s_1,x_0)\partial_xF(s_2,x_0)\partial_xF(s_3,x_0)
\ee
which agrees with the derivation based on the Picard iteration. It is straightforward but cumbersome to verify that Eq.~\eqref{ax:result} also holds for larger $n$.

\subsubsection{Order-by-order evaluation}

We consider the terms on the RHS of Eq.~\eqref{ax:identity} as a series in orders of $\mu$ and establish the equality of each terms on both sides of Eq.~\eqref{ax:identity}. Considering orders up to $\mu^3$ we thus need to proof the following equalities:
\be
\mu:\qquad&& i\int_0^t\D s\left<e^{i\int_{0}^t\D u\, z(u)\dot{X}(u)}g(s)F(s,X(s))\right>_0=i\int_0^t\D s\,z(s)\phi_1(s,x_0)\label{ax:term1}\\
\mu^2:\qquad&& -\frac{1}{2}\int_0^t\D s_1\int_0^t\D s_2\left<e^{i\int_{0}^t\D u\, z(u)\dot{X}(u)}g(s_1)g(s_2)F(s_1,X(s_1))F(s_2,X(s_2))\right>_0\nonumber\\
&&=i\int_0^t\D s\,z(s)\phi_2(s,x_0)-\frac{1}{2}\int_0^t\D s_1\int_0^t\D s_2\,z(s_1)z(s_2)\phi_1(s_1,x_0)\phi_1(s_2,x_0)\label{ax:term2}\\
\mu^3:\qquad&& -\frac{i}{6}\int_0^t\D s_1\int_0^t\D s_2\int_0^t\D s_3\left<e^{i\int_{0}^t\D u\, z(u)\dot{X}(u)}g(s_1)g(s_2)g(s_3)\right.\times\nonumber\\
&&\left.F(s_1,X(s_1))F(s_2,X(s_2))F(s_3,X(s_3))\vphantom{e^{i\int_{0}^t\D u\, z(u)\dot{X}(u)}}\right>_0\nonumber\\
&&=i\int_0^t\D s\,z(s)\phi_3(s,x_0)-\int_0^t\D s_1\int_0^t\D s_2\,z(s_1)z(s_2)\phi_1(s_1,x_0)\phi_2(s_2,x_0)\nonumber\\
&&\quad-\frac{i}{6}\int_0^t\D s_1\int_0^t\D s_2\int_0^t\D s_3\,z(s_1)z(s_2)z(s_3)\phi_1(s_1,x_0)\phi_1(s_2,x_0)\phi_1(s_3,x_0)\label{ax:term3}
\ee

Considering Eq.~\eqref{ax:term1}, we have
\be
i\int_0^t\D s\left<e^{i\int_{0}^t\D u\, z(u)\dot{X}(u)}g(s)F(s,X(s))\right>_0&=&\int_0^t\D s\left<\frac{\delta}{\delta \dot{X}(s)}e^{i\int_{0}^t\D u\, z(u)\dot{X}(u)}F(s,X(s))\right>_0\nonumber\\
&=&i\int_0^t\D s\,z(s)F(s,x_0)\nonumber\\
&=&i\int_0^t\D s\,z(s)\phi_1(s,x_0)
\ee
using partial integration on the path weight in the first step and subsequently Eqs.~(\ref{ax:cancel},\ref{ax:phi0}).

Considering Eq.~\eqref{ax:term2}, we have
\be
&&-\frac{1}{2}\int_0^t\D s_1\int_0^t\D s_2\left<e^{i\int_{0}^t\D u\, z(u)\dot{X}(u)}g(s_1)g(s_2)F(s_1,X(s_1))F(s_2,X(s_2))\right>_0\nonumber\\
&&=\frac{1}{2}\int_0^t\D s_1\int_0^t\D s_2\left<\frac{\delta}{\delta \dot{X}(s_1)}\frac{\delta}{\delta \dot{X}(s_2)}e^{i\int_{0}^t\D u\, z(u)\dot{X}(u)}F(s_1,X(s_1))F(s_2,X(s_2))\right>_0
\ee
The combinatorics of the higher order partial derivatives becomes quickly tedious. Here, the situation is simplified since the $s_{1,2}$ variables are indistinguishable and we can use the general formula
\be
\label{ax:multi}
\frac{\delta^n}{\delta \dot{X}(s_1)\cdots \delta \dot{X}(s_n)}h_{\rm a}(s_1,...,s_n)h_{\rm b}(s_1,...,s_n)=\sum_{l=0}^n\left(\begin{matrix}n \\ l\end{matrix}\right)\delta^lh_{\rm a}\,\delta^{n-l}h_{\rm b},
\ee
where the shorthand notation $\delta^lh_{\rm a}=\frac{\delta^l}{\delta \dot{X}(s_1)\cdots \delta \dot{X}(s_l)}h_{\rm a}(s_1,...,s_l)$ is used with $\delta^0h_{\rm a}=h_{\rm a}$. This yields
\be
&&\frac{1}{2}\int_0^t\D s_1\int_0^t\D s_2\left<\frac{\delta}{\delta \dot{X}(s_1)}\frac{\delta}{\delta \dot{X}(s_2)}e^{i\int_{0}^t\D u\, z(u)\dot{X}(u)}F(s_1,X(s_1))F(s_2,X(s_2))\right>_0\nonumber\\
&&=\frac{1}{2}\int_0^t\D s_1\int_0^t\D s_2\left\{(-1)z(s_1)z(s_2)\left<\vphantom{\frac{\delta}{\delta \dot{X}(s_1)}}F(s_1,X(s_1))F(s_2,X(s_2))\right>_0\right.\nonumber\\
&&\quad+2i\,z(s_2)\left<\frac{\delta}{\delta \dot{X}(s_1)}F(s_1,X(s_1))F(s_2,X(s_2))\right>_0\nonumber\\
&&\quad+\left.\left<\frac{\delta}{\delta \dot{X}(s_1)}\frac{\delta}{\delta \dot{X}(s_2)}F(s_1,X(s_1))F(s_2,X(s_2))\right>_0\right\}\nonumber\\
&&=-\frac{1}{2}\int_0^t\D s_1\int_0^t\D s_2\,z(s_1)z(s_2)F(s_1,x_0)F(s_2,x_0)+i\int_0^t\D s\,z(s)\phi_2(s,x_0),
\ee
confirming Eq.~\eqref{ax:term2}. In the first step, we have used the factorization property of the average $\left<f_{\rm a}(X(s_1))f_{\rm b}(X(s_2))\right>_0=f_{\rm a}(x_0)\left<f_{\rm b}(X(s_2))\right>_0$. In the last step, we have used Eq.~\eqref{ax:result} with $n=2$ and Eq.~\eqref{ax:cancel}, which leads to the vanishing of the third term.

Considering Eq.~\eqref{ax:term3}, we use again Eq.~\eqref{ax:multi} to calculate the functional derivatives
\be
\qquad&& -\frac{i}{6}\int_0^t\D s_1\int_0^t\D s_2\int_0^t\D s_3\left<e^{i\int_{0}^t\D u\, z(u)\dot{X}(u)}g(s_1)g(s_2)g(s_3)\right.\times\nonumber\\
&&\left.F(s_1,X(s_1))F(s_2,X(s_2))F(s_3,X(s_3))\vphantom{e^{i\int_{0}^t\D u\, z(u)\dot{X}(u)}}\right>_0\nonumber\\
&&=\frac{1}{6}\int_0^t\D s_1\int_0^t\D s_2\int_0^t\D s_3\left<\frac{\delta}{\delta \dot{X}(s_1)}\frac{\delta}{\delta \dot{X}(s_2)}\frac{\delta}{\delta \dot{X}(s_3)}e^{i\int_{0}^t\D u\, z(u)\dot{X}(u)}\right.\times\nonumber\\
&&\quad\left.\vphantom{\frac{\delta}{\delta \dot{X}(s_2)}}F(s_1,X(s_1))F(s_2,X(s_2))F(s_3,X(s_3))\right>_0\nonumber\\
&&=\frac{1}{6}\int_0^t\D s_1\int_0^t\D s_2\int_0^t\D s_3\left\{\vphantom{\frac{\delta}{\delta \dot{X}(s_2)}}(-i)z(s_1)z(s_2)z(s_3)F(s_1,x_0)F(s_2,x_0)F(s_3,x_0)\right.\nonumber\\
&&\quad-3\,z(s_2)z(s_3)\left<\frac{\delta}{\delta \dot{X}(s_1)}F(s_1,X(s_1))F(s_2,X(s_2))F(s_3,X(s_3))\right>_0\nonumber\\
&&\quad+\left.3iz(s_3)\left<\frac{\delta}{\delta \dot{X}(s_1)}\frac{\delta}{\delta \dot{X}(s_2)}F(s_1,X(s_1))F(s_2,X(s_2))F(s_3,X(s_3)\right>_0\right\}\nonumber\\
&&=\frac{1}{6}\int_0^t\D s_1\int_0^t\D s_2\int_0^t\D s_3\left\{\vphantom{\frac{\delta}{\delta \dot{X}(s_2)}}(-i)z(s_1)z(s_2)z(s_3)\phi_1(s_1,x_0)\phi_1(s_2,x_0)\phi_1(s_3,x_0)\right.\nonumber\\
&&\quad-6\,z(s_2)z(s_3)F(s_1,x_0)\Theta(s_2-s_1)\partial_xF(s_2,x_0)F(s_3,x_0)+\left.6iz(s_3)\phi_3(s_3,x_0)\vphantom{\frac{\delta}{\delta \dot{X}(s_2)}}\right\}
\ee
using again Eqs.~(\ref{ax:result},\ref{ax:cancel}) and thus confirming Eq.~\eqref{ax:term3}. The correspondence for higher orders can be shown following the same calculation techniques.

\section{Formalism for velocity-dependent forces}
\label{Sec:vecforce}

When $\vec{F}$ includes a velocity-dependence as in the hydrodynamic force Eq.~\eqref{ax:hydrof} below, it is necessary to extend the microscopic density field to include velocity degrees of freedom
\be
\rho(\vec{x},\vec{v},t)=\sum_{i=1}^N\delta(\vec{x}-\vec{Y}_i(t))\delta(\vec{v}-\dot{\vec{Y}}_i(t)),
\ee
which leads to the equation of motion
\be
\label{eqm2}
\dot{\vec{X}}(t)=\mu\int\D\vec{x}\int\D\vec{v}\,\vec{F}(\vec{x}-\vec{X}(t),\vec{v})\rho(\vec{x},\vec{v},t).
\ee
The further derivation follows the same steps as in the main text. The main ingredient is the $n$-point function to first order in $\rho_0$, which now reads
\be
\label{ax:npointv}
\left<\rho(\vec{x}_1,\vec{v}_1,t_1)\cdots\rho(\vec{x}_n,\vec{v}_n,t_n)\right>&=&\rho_0\int\D\vec{y}_0\left<\delta(\vec{x}_1-\vec{Y}(t_1))\delta(\vec{v}_1-\dot{\vec{Y}}(t_1))\cdots\right.\nonumber\\
&&\left.\cdots\delta(\vec{x}_n-\vec{Y}(t_n))\delta(\vec{v}_n-\dot{\vec{Y}}(t_n))\right>_{\vec{Y}(0)=\vec{y}_0}.
\ee
Eq.~\eqref{ax:npointv} can be shown following a similar calculation as in Appendix~Sec.~\ref{Sec:npoint}. The proof of Eq.~\eqref{series} is then unchanged, see Appendix~Sec.~\ref{Sec:proof}. This can be easily seen in the heuristic proof, but also in the order-by-order calculation, since the additional $\dot{\vec{Y}}$-dependence in the force is simply contained in the explicit $t$-dependence in Eq.~\eqref{ax:ode}. The solution of the two-body problem again follows from the Picard iteration of the ODE, which is now
\be
\vec{X}(t)=\mu\int_0^t\D s\,\vec{F}(\vec{Y}(s)-\vec{X}(s),\dot{\vec{Y}}(s)),
\ee
with $\vec{X}(0)=\vec{0}$. Up to second order we have then
\be
\vec{X}^{s,2}(t)&=&\mu\int_0^t\D s\,\vec{F}\left(\vec{Y}(s)-\mu\int_0^s\D s'\,\vec{F}\left(\vec{Y}(s'),\dot{\vec{Y}}(s')\right),\dot{\vec{Y}}(s)\right)
\ee
and thus
\be
\vec{f}_\mu[\vec{Y}]\approx\mu\,\vec{F}\left(\vec{Y}(t)-\mu\int_0^t\D s'\,\vec{F}\left(\vec{Y}(s'),\dot{\vec{Y}}(s')\right),\dot{\vec{Y}}(s)\right).
\ee
Taylor expansion in orders of $\mu$ then yields the approximations
\be
\vec{f}_{\mu^1}[\vec{Y}]&=&\mu\,\vec{F}(\vec{Y}(t),\dot{\vec{Y}}(t))\label{ax:fapprox1}\\
\vec{f}_{\mu^2}[\vec{Y}]&=&\mu\,\vec{F}(\vec{Y}(t),\dot{\vec{Y}}(t))+\mu^2\nabla_{\vec{x}}\vec{F}(\vec{Y}(t),\dot{\vec{Y}}(t))\int_0^t\D s\,\vec{F}(\vec{Y}(s),\dot{\vec{Y}}(s)).\label{ax:fapprox2}
\ee

\section{The characteristic functional of the spatial Poisson process}
\label{Sec:cf}

We first assume a finite volume $\tilde{V}$. Substituting Eq.~\eqref{spatialP} into Eq.~\eqref{cf} yields
\be
&&\left<\exp\left\{i\int_{0}^t \vec{k}(s)\cdot\vec{\dot{X}}(s)\D s\right\}\right>\nonumber\\
&&=\left<\exp\left\{i\int_{0}^t \vec{k}(s)\cdot\sum_{\vec{y}_0\in \Phi}\vec{f}_\mu[\vec{Y}(s)]\D s\right\}\right>\nonumber\\
&&=\sum_{n=0}^\infty P(\tilde{N}=m)\prod_{i=0}^m\frac{1}{\tilde{V}}\int_{\tilde{V}}\D\vec{y}_0\left<\exp\left\{i\int_{0}^t \vec{k}(s)\cdot\vec{f}_\mu[\vec{Y}(s)]\D s\right\}\right>_{\vec{Y}(0)=\vec{y}_0}\nonumber\\
&&=\sum_{m=0}^\infty \frac{(\rho_0\tilde{V})^m}{m!}e^{-\rho_0\tilde{V}}\left(\frac{1}{\tilde{V}}\int_{\tilde{V}}\D\vec{y}_0\left<\exp\left\{i\int_{0}^t \vec{k}(s)\cdot\vec{f}_\mu[\vec{Y}(s)]\D s\right\}\right>_{\vec{Y}(0)=\vec{y}_0}\right)^m\nonumber\\
&&=\exp\left(-\rho_0\tilde{V}+\rho_0\int_{\tilde{V}}\D\vec{y}_0\left<\exp\left\{i\int_{0}^t \vec{k}(s)\cdot\vec{f}_\mu[\vec{Y}(s)]\D s\right\}\right>_{\vec{Y}(0)=\vec{y}_0}\right)\label{ax:cfP}
\ee
using the independence, uniformity, and Poisson statistics of of the spatial Poisson process $\Phi$. Upon taking the limit $\tilde{V}\to\infty$, Eq.~\eqref{ax:cfP} just becomes Eq.~\eqref{main}.

\section{Mapping the spatial Poisson process to a non-Markovian Poisson process in time}
\label{Sec:mapping}

In this section I consider swimmers moving in straight lines according to
\be
\label{ax:straight}
\vec{Y}(t)=\vec{y}_0+v_{\rm A}\vec{\hat{n}}t,
\ee
with a constant swimming speed $v_{\rm A}$. The initial position $\vec{y}_0$ and swimming direction $\vec{\hat{n}}$ are chosen uniformly and isotropically distributed, respectively. Using this specific swimmer model and the assumption of independent scattering events between the swimmers and the tracer, a coloured Poisson process has been derived for the tracer dynamics in \cite{Kanazawa:2020aa}, which specifies the characteristic functional Eq.~\eqref{cf} as
\be
\label{ax:cfold}
\ln \psi_\vec{\dot{X}}[\vec{k}]=\int_{-\infty}^\infty\D t'\int\D\vec{b}\int_{-\pi}^\pi\D\phi'\,\lambda(\vec{b})\left(\exp\left\{i\int_{0}^t \D s\,\vec{k}(s)\cdot\vec{f}^{\rm O}_{\vec{b},\phi'}(s-t')\right\}-1\right)
\ee
where $\vec{f}^{\rm O}_{\vec{b},\phi'}$ is the force shape function due to a single swimmer--tracer scattering event. The intensity of the Poisson process is given by $\lambda(\vec{b})=\frac{\rho_0v_{\rm A}}{4\pi\,b}$, which depends on the impact parameter vector $\vec{b}$. The force shape function $\vec{f}^{\rm O}_{\vec{b},\phi'}$ depends also on the injection angle $\phi'$ of the swimmer in the plane normal to $\vec{b}$ and has been derived in closed analytical form from the two-body interaction Eq.~\eqref{ax:2body} with the hydrodynamic force \cite{Leptos:2009aa}
\be
\label{ax:hydrof}
\vec{F}_{\rm hyd}(\vec{x},\vec{\hat{n}})=\frac{p}{\vec{x}^2}\left(3\frac{(\vec{\hat{n}}\cdot\vec{x})^2}{\vec{x}^2}-1\right)\frac{\vec{x}}{|\vec{x}|},
\ee
where the parameter $p$ specifies the strength of the dipole force. The non-Markovian process Eq.~\eqref{ax:cfold} reproduces the dynamical features of the tracer dynamics observed in simulations and experiments, predicting in particular a L\'evy flight regime underlying the observed enhancement of the tracer diffusion \cite{Kanazawa:2020aa}. In the following it is shown that these results follow from Eq.~\eqref{main} for the special case of the swimmer dynamics given by Eq.~\eqref{ax:straight}.

\begin{figure}
\centering
\includegraphics[height=4cm]{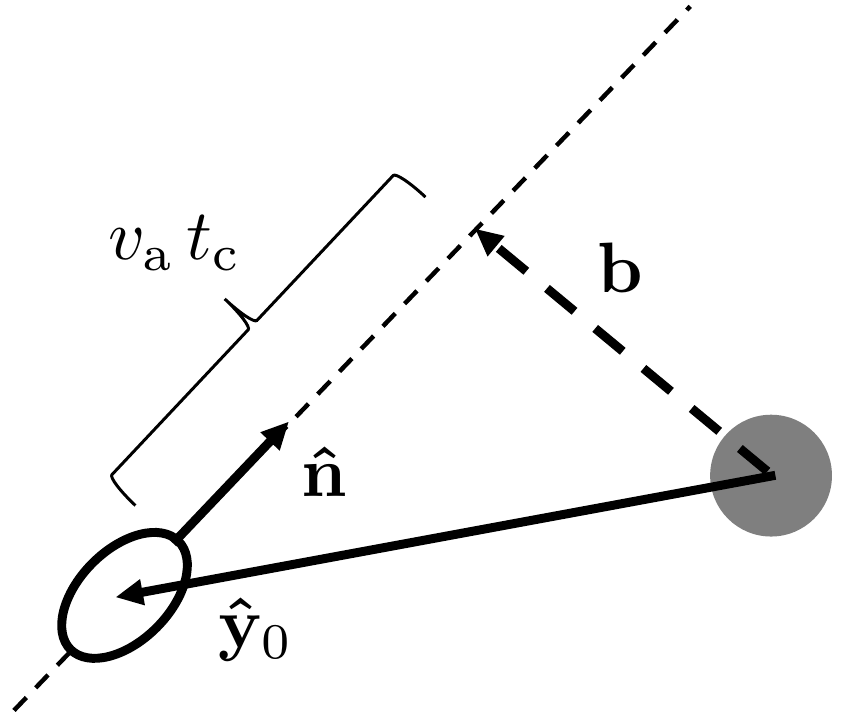}
\caption{\label{Fig_trans} Illustration of the coordinate transformation in two dimensions. The swimmer (ellipse) is moving in the direction $\vec{\hat{n}}$ starting from the initial position $\vec{y}_0$ relative to the tracer (solid disk) at the origin. The new coordinates describe the swimmer motion with the impact parameter vector $\vec{b}$, which satisfies $\vec{b}\cdot\vec{\hat{n}}=0$. The remaining degree of freedom in the new coordinates is $\phi'$ describing the angle of the swimmer direction in the plane normal to $\vec{b}$ (not indicated), see Fig.~1c of \cite{Kanazawa:2020aa}.}
\end{figure}

Starting with the first order approximation of $\vec{f}_{\mu^1}$, Eq.~\eqref{ax:fapprox1}, we obtain
\be
\vec{f}_{\vec{\mu}}[\vec{Y}(t)]&\approx& \mu\,\vec{F}(\vec{Y}(t),\dot{\vec{Y}}(t))\nonumber\\
&=&\mu\,\vec{F}_{\rm hyd}(\vec{y}_0+v_{\rm A}\vec{\hat{n}}t,\vec{\hat{n}})\nonumber\\
&=&\frac{\mu\,p}{(\vec{y}_0+v_{\rm A}\vec{\hat{n}}t)^2}\left(3\frac{((\vec{y}_0+v_{\rm A}\vec{\hat{n}}t)\cdot\vec{\hat{n}})^2}{(\vec{y}_0+v_{\rm A}\vec{\hat{n}}t)^2}-1\right)\frac{(\vec{y}_0+v_{\rm A}\vec{\hat{n}}t)}{\sqrt{(\vec{y}_0+v_{\rm A}\vec{\hat{n}}t)^2}}
\ee
using Eqs.~(\ref{ax:straight},\ref{ax:hydrof}). Writing out the average in Eq.~\eqref{main} yields
\be
\label{ax:step1}
\ln \psi_\vec{\dot{X}}[\vec{k}]=\frac{\rho_0}{4\pi}\int\D \vec{y}_0\oint\D\vec{\hat{n}}\left(\exp\left\{i\mu\int_{0}^t \D s\,\vec{k}(s)\vec{F}_{\rm hyd}(\vec{y}_0+v_{\rm A}\vec{\hat{n}}t,\vec{\hat{n}})\right\}-1\right)
\ee
The remaining integrals capture integrations over five degrees of freedom of the swimmer motion, which is the same dimensionality as for the integrals of Eq.~\eqref{ax:cfold}. We now implement a variable transformation between these sets of variables, which is illustrated in Fig.~\ref{Fig_trans} in two dimensions. We emphasize that the impact parameter vector $\vec{b}$ is constrained to be perpendicular to the swimmer direction $\vec{\hat{n}}$, see Fig.~\ref{Fig_trans}, thus there is only one degree of freedom ($\phi'$) describing the swimmer direction in addition to $\vec{b}$. The vector $\vec{b}$ is thus given as
\be
\label{ax:bvec}
\vec{b}=\vec{y}_0+v_{\rm A}\vec{\hat{n}}t_{\rm c},\qquad\qquad t_{\rm c}=-\vec{y}_0\cdot\vec{\hat{n}}/v_{\rm A},
\ee 
where $t_{\rm c}$ is the time needed such that the swimmer starting at $\vec{y}_0$ satisfies $\vec{b}\cdot\vec{\hat{n}}=0$. We obtain further $\vec{y}_0+v_{\rm A}\vec{\hat{n}}t=\vec{b}+v_{\rm A}\vec{\hat{n}}(t-t_{\rm c})$ and thus in Eq.~\eqref{ax:step1}
\be
\label{ax:offset}
\vec{F}_{\rm hyd}(\vec{y}_0+v_{\rm A}\vec{\hat{n}}t,\vec{\hat{n}})&=&\vec{F}_{\rm hyd}(\vec{b}+v_{\rm A}\vec{\hat{n}}(t-t_{\rm c}),\vec{\hat{n}})
\ee
We now show that $\vec{F}_{\rm hyd}(\vec{b}+v_{\rm A}\vec{\hat{n}}t,\vec{\hat{n}})$ is already the first order approximation $f^{\rm O(1)}_{\vec{b},\phi'}(t)$ of the force shape function calculated in \cite{Kanazawa:2020aa}, which can be seen after expressing $\vec{b},\vec{\hat{n}}$ in spherical coordinates. To this end we introduce a spherical coordinate system $\vec{b}=(b\sin \theta\cos\phi,b\sin\theta\sin\phi,b\cos\theta)^{\rm T}$ with orthogonal unit vectors
\be
\vec{\hat{e}}_b=\left(\begin{matrix} \sin\theta\cos\phi\\ \sin\theta\sin\phi \\ \cos\theta\end{matrix}\right),\quad \vec{\hat{e}}_\theta=\left(\begin{matrix} \cos\theta\cos\phi\\ \cos\theta\sin\phi \\ -\sin\theta\end{matrix}\right),\quad \vec{\hat{e}}_\phi=\left(\begin{matrix} -\sin\phi\\ \cos\phi \\ 0\end{matrix}\right)
\ee
Expressing $\vec{\hat{n}}$ in terms another set of spherical coordinates in this system via $\vec{\hat{n}}=\sin \theta'\cos\phi'\vec{\hat{e}}_\theta+\sin \theta'\sin\phi'\vec{\hat{e}}_\phi+\cos\theta'\vec{\hat{e}}_b$ \cite{Kanazawa:2020aa}, we see that the constraint $\vec{b}\cdot\vec{\hat{n}}=b\vec{\hat{e}}_b\cdot\vec{\hat{n}}=0$ imposes $\cos\theta'=0$ and thus $\sin\theta'=1$ which yields
\be
\label{ax:nhat}
\vec{\hat{n}}=\left(\begin{matrix} \cos\theta\cos\phi\cos\phi'-\sin\phi\sin\phi' \\ \cos\theta\sin\phi\cos\phi' +\cos\phi\sin\phi' \\ -\sin\theta\cos\phi'\end{matrix}\right)
\ee
Noting that $(\vec{b}+v_{\rm A}\vec{\hat{n}}t)^2=b^2+v_{\rm A}^2t^2$ we obtain
\be
\label{ax:fs1}
\vec{F}_{\rm hyd}(\vec{b}+v_{\rm A}\vec{\hat{n}}t,\vec{\hat{n}})=\frac{p}{b^2+v_{\rm A}^2t^2}\left(3\frac{v_{\rm A}^2t^2}{b^2+v_{\rm A}^2t^2}-1\right)\frac{(\vec{b}+v_{\rm A}\vec{\hat{n}}t)}{\sqrt{b^2+v_{\rm A}^2t^2}}
\ee
which together with Eq.~\eqref{ax:nhat} is just Eq.~(S69) in \cite{Kanazawa:2020aa}. The second order contribution $\mu^2\nabla_{\vec{x}}\vec{F}(\vec{Y},\dot{\vec{Y}})\int_0^t\D s\vec{F}(\vec{Y}(s),\dot{\vec{Y}}(s))$ in $\vec{f}_{\mu^2}$ of Eq.~\eqref{ax:fapprox2} can now be calculated in the coordinates $(\vec{b},\phi')$ by integrating Eq.~\eqref{ax:fs1} with respect to $t$, which yields a closed analytical result. This is equivalent to the method used in \cite{Kanazawa:2020aa} to calculate the second order contribution to the force shape function $f^{\rm O(2)}_{\vec{b},\phi'}(t)$ and thus gives the same result.

The remaining step is to calculate the Jacobian of the transformation between the set of variables $(\vec{y}_0,\vec{\hat{n}})$ used in Eq.~\eqref{ax:step1} and the variables $(t_{\rm c},\vec{b},\phi')$ used in Eq.~\eqref{ax:cfold} ($t_{\rm c}$ is just the offset $t'$ in Eq.~\eqref{ax:offset}). This is done easiest by keeping Cartesian coordinates for $(\vec{y}_0,\vec{\hat{n}})$, whereby the orientational integral $\oint\D\vec{\hat{n}}$ can be expressed as a volume integral over a delta function enforcing the unit sphere surface
\be
\frac{\rho_0}{4\pi}\int\D \vec{y}_0\oint\D\vec{\hat{n}}\left(\vphantom{\exp\int}...\right)&=&\frac{\rho_0}{4\pi}\int\D \vec{y}_0\int\D n_x\int\D n_y\int\D n_z\delta\left(\sqrt{n_x^2+n_y^2+n_z^2}-1\right)\left(\vphantom{\exp\int}...\right)\nonumber\\
&=&\frac{\rho_0}{4\pi}\int\D \vec{y}_0\int_{-1}^1\D n_x\int_{-\sqrt{1-n_x^2}}^{\sqrt{1-n_x^2}}\D n_y\frac{2}{\sqrt{1-n_x^2-n_y^2}}\left(\vphantom{\exp\int}...\right)
\ee
The equations for the transformation are then (see Eqs.~(\ref{ax:bvec},\ref{ax:nhat}))
\be
\vec{y}_0&=&\vec{b}-v_{\rm A}\vec{\hat{n}}t_{\rm c}\\
n_x&=&\cos\theta\cos\phi\cos\phi'-\sin\phi\sin\phi'\\
n_y&=&\cos\theta\sin\phi\cos\phi' +\cos\phi\sin\phi'
\ee
which gives the Jacobian determinant
\be
|\mathcal{J}|=v_{\rm A}b\sin^2\theta \cos\phi'.
\ee
In addition, we have in the new coordinates
\be
\sqrt{1-n_x^2-n_y^2}=\sin\theta\cos\phi'.
\ee
The remaining step is to find the correct boundaries in the new set of coordinates. To this end we note that $(\vec{y}_0,\vec{\hat{n}})$ for a given swimmer is uniquely mapped onto a set $(t_{\rm c},\vec{b},\phi')$ with $t_{\rm c}\in [0,\infty)$ and $\phi'\in[-\pi,\pi]$. Overall, we obtain then the transformation
\be
\frac{\rho_0}{4\pi}\int\D \vec{y}_0\oint\D\vec{\hat{n}}\left(\vphantom{\exp\int}...\right)&=&\int_0^\infty\D t_{\rm c}\int_0^\infty\D b\int_0^\pi\D\theta\int_{-\pi}^\pi\D\phi\int_{-\pi}^\pi\D\phi'\,\frac{\rho_0v_{\rm A}}{2\pi}b\sin\theta\left(\vphantom{\exp\int}...\right)\nonumber\\
&=&\int_{-\infty}^\infty\D t_{\rm c}\int\D \vec{b}\int_{-\pi}^\pi\D\phi'\,\frac{\rho_0v_{\rm A}}{4\pi\,b}\left(\vphantom{\exp\int}...\right)\nonumber\\
\ee
where we recover the intensity $\lambda(\vec{b})$ of the coloured Poisson process as in Eq.~\eqref{ax:cfold}.

\end{appendix}

\end{document}